\documentstyle[12pt,aaspp4]{article}

\def\MJ {$M_{J}$}

\def\ms {m s$^{-1}$}

\def\ups{$\upsilon$ And}
\def\degr{$^{\circ}$}

\received{May 17th, 1999}
\accepted{July 9th, 1999}
\slugcomment{Accepted for publication in The Astrophysical Journal Letters}

\begin{document}

\title{ Analysis of the Hipparcos Measurements of $\upsilon$ Andromedae ---\\
A Mass Estimate of its Outermost Known Planetary Companion}


\author{Tsevi Mazeh and Shay Zucker}
\affil{School of Physics and Astronomy, Raymond and Beverly Sackler
Faculty of Exact Sciences, Tel Aviv University, Tel Aviv, Israel\\
mazeh@wise7.tau.ac.il; shay@wise.tau.ac.il} 
\author{Andrea Dalla Torre\footnotemark and Floor van Leeuwen}
\affil{Institute of Astronomy, Cambridge University, Cambridge, UK\\
fvl@ast.cam.ac.uk; torre@ast.cam.ac.uk}

\footnotetext{ On leave from the University of Padova}
\vskip 3. truecm

\begin{abstract}
We present an analysis of Hipparcos astrometric measurements of \ups, a
nearby main-sequence star around which three planet candidates have
recently been discovered by means of radial-velocity measurements. The
stellar orbit associated with the outermost candidate has a period of
$1269\pm 9$ days and a minimum semi-major axis of 0.6 milli-arc-sec
(mas).  Using the Hipparcos data {\it together} with the spectroscopic
elements we found a semi-major axis of $1.4\pm0.6$ mas.  This implies
a mass of $10.1^{+4.7}_{-4.6}$ Jupiter masses for that 
planet of \ups.

\subjectheadings{astrometry --- planetary systems --- stars: individual
($\upsilon$ Andromedae)}
\end{abstract}
\newpage

\section{INTRODUCTION}

About twenty candidates for extrasolar planets have been announced over
the past four years (e.g., Mayor \& Queloz 1995; Noyes et al.\ 1997;
Marcy \& Butler 1998).  In each case, very precise stellar
radial-velocity measurements, with a precision of 10~m~s$^{-1}$ or
better, indicated the presence of a low-mass unseen companion orbiting a
nearby solar-type star. These high-precision discoveries came almost a
decade after the first 'planet candidate' around HD 114762 was
discovered (Latham et al.\ 1989; Mazeh, Latham \& Stefanik 1996) with
much lower precision. In all cases, the individual masses of the
companions are not known, because the inclination angles of their
orbital planes relative to our line of sight cannot be derived from the
spectroscopic data. The minimum masses for all candidates, attained for
an inclination angle of $90^{\circ}$, are in the range 0.5 to 10 Jupiter
masses (\MJ).

Although the only feature that characterized the 'planet candidates' was
the smallness of their mass, no {\it direct} mass estimates were
available.  The near consensus on their nature as planets (e.g., Boss
1996; Mazeh, Goldberg \& Latham 1998; Marcy \& Butler 1998; but see
Black 1997) was based only on statistical arguments. It relied on the
fact that for orbital planes randomly oriented in space, the expectation
value of the sine of the inclination angle is close to unity.  Still,
the mass estimation of individual planet candidates is extremely
important. First, we want to know what is the actual mass range of the
newly discovered objects. Some of the known candidates might have low
inclinations and therefore substantially larger masses than their
minimum masses. Second, the growing population of known extra-solar
planets is starting to reveal interesting features, like non-zero
eccentricity (e.g., Marcy \& Butler 1998; see also Mazeh, Mayor \&
Latham 1996) and unexpectedly close orbits (e.g., Marcy, Cochran \&
Mayor 2000). To find whether and how these features depend on the planet
mass we need better estimates of their masses. The spectroscopic data
cannot yield any information beyond the {\it minimum} mass. We need
additional information, like precise astrometry of the orbit, from which
we can derive the inclination, and therefore the secondary mass, at
least for the cases where the primary mass can be estimated from its
spectral type.

At present, the astronomical community has at hand the accurate
astrometric Hipparcos data, which have already yielded numerous orbits
with small semi-major axes (ESA 1997; S\"oderhjelm 1999).  The newly
discovered planets are expected to induce reflex motion on their primary
stars with semi-major axes of the order of 1~milli-arc-sec (mas) or
smaller, below the detection threshold of the satellite.  However, the
spectroscopic elements yield only lower limit to the semi-major axis.
The actual semi-major axis of the astrometric orbit is inversely
proportional to the sine of the inclination angle, and therefore can be
large enough to be detectable by the Hipparcos observations.
Furthermore, the use of the radial-velocity orbital elements, the
orbital period in particular, can improve our astrometric detection
limit significantly.  Moreover, even if the orbit is too small for
definite detection by Hipparcos, any upper limit of the astrometric
orbit which is not substantially larger than the minimum semi-major axis
has its own astrophysical significance.

Perryman et al. (1996) applied this approach to 47 UMa, 70 Vir and 51
Peg, using the Hipparcos data to put upper limits on the masses of their
unseen companions. Their 90\% confidence limit for the three
planet-candidates were 22, 65, and 1100 \MJ, respectively.  Another
opportunity to utilize this approach came recently, when Butler et al.\ 
(1999) announced their discovery of three planets around \ups\ (=HD
9826=HR 458=HIP 7513). The most distant known planet candidate of that
system induces stellar orbital motion with semi-major axis that can be
no smaller than about 0.6~mas --- a lower limit not terribly smaller
than the detection threshold of Hipparcos. Hipparcos measured the
(one-dimensional) astrometric position of \ups\ on 27 independent
reference circles, each circle providing two partially independent
reduction results (see, for example, Perryman et al. 1997; van Leeuwen
1997).  We hoped to get a relatively small upper limit for the
semi-major axis, because the goodness-of-fit statistic given for \ups\ 
in the Hipparcos catalogue (F2 in column 30) has a value of 0.08, which
does not indicate the presence of any unresolved orbital motion.

Our analysis is based on a method described by van Leeuwen
and Evans (1998), a technique which was further developed for the use on
astrometric orbits by two of us --- ADT and FvL. The method refers back
to the intermediate astrometric data (the abscissa residuals), which are
fitted with a model of the apparent motion of the star through the
optimization of a number of parameters (Dalla Torre and van Leeuwen, in
prep).  With these advanced methods at hand, we set out to reanalyze the
54 Hipparcos data points of \ups. This paper reports on our results.

\section{The Hipparcos Astrometry}  

The analysis used the spectroscopic elements of the
``updated Lick data orbit'' as presented by Laughlin \& Adams (1999, see
also Butler et al. 1999). These included the
spectroscopic period, $P=1269\pm8.5$ day, the periastron passage,
$T_0=2453813\pm32$, the radial-velocity amplitude, $K=69.5\pm2.1$ \ms,
and eccentricity $e=0.30\pm0.05$, and the longitude of the periastron,
$\omega=236\pm15$. The orbital astrometric elements include $P, \, T_0,
\, e, \, \omega$ and three additional elements --- the semi-major axis,
$a_1$, the inclination, $i$, and the longitude of the nodes,
$\Omega$. In addition, the astrometric solution includes the five regular
astrometric parameters --- the parallax, the position (in right ascension
and declination) and the proper motion (in right ascension and
declination). All together we had 12 parameter model to fit to the
astrometric data, four of which are common with the spectroscopic orbit.

To find the best astrometric orbit we used the values of $P, T_0, e,
\omega$ as given by the spectroscopic orbit, and solved for the other
parameters. To do that we considered a dense grid on the ($a_1, i$) plane,
and found the values of the five regular astrometric parameters and
$\Omega$ that minimized the $\chi^2$ statistics for each value of ($a_1,
i$).  The result of this search is a $\chi^2$ {\it function}, which
depends on $a_1$ and $i$. The square-root normalized $\chi^2$ is plotted
in Figure~1 as a two-dimensional function.

To our surprise we found a very pronounced ``valley'' at about 1.4 mas,
indicating a detection of an astrometric motion. The valley
runs across all possible inclinations, indicating that the analysis so
far could derive the semi-major axis but not the inclination. This is
probably due to the small amplitude of the astrometric motion and
because the Hipparcos satellite did not measure any two-dimensional
stellar position, but only a projected stellar position on the
instantaneous reference circle of the satellite.

To derive the best inclination for the system we used another
spectroscopic element --- the radial-velocity amplitude $K$, which has
not been used so far in the analysis. This element induces a constraint
on the product of $a_1$ and $\sin i$, which for the \ups\ case results
in 

\begin{equation}
a_1 \times \sin i= 0.56\pm0.02\
            \Bigl({{P}\over{1269\,{\rm day}}}\Bigr)
            \Bigl({{K}\over{69.5\,{\rm m\,s^{-1}}}}\Bigr)
            \Bigl({{\sqrt{1-e^2}}\over{0.95}}\Bigr)
            \Bigl({{\pi}\over{74.25\,{\rm mas}}}\Bigr)
            \ {\rm mas} \ ,
\end{equation}
where we have used here the Hipparcos {\it catalog} parallax,
$\pi=74.25\pm0.72$ mas. This constraint is plotted in Figure 1 as a
continuous line both on the ($a_1, i$) plane and on the square-root
normalized $\chi^2$ surface. In Figure 2 we collapsed the
two-dimensional function onto the line of
Eq. (1). We got a clear minimum at $156\fdg0$. This corresponded to a
semi-major axis of 1.4 mas and a mass of 10.1 \MJ. From Figure 2 we could
also derive a $1\, \sigma$ range of inclinations --- [$131\fdg4,
163\fdg9$] which resulted in

\begin{equation}
a_1= 1.4 \pm 0.6\,{\rm mas} \ \ ,\,\ M_p=10.1^{+4.7}_{-4.6} \,M_J \ ,
\end{equation}
where $M_p$ is the mass of the outermost planet. We give the upper and
the lower limit of the planet mass to emphasize the non-symmetric
distribution of the errors. This is more pronounced when we move to the 
$2\, \sigma$ range, [$14\fdg1,\,167\fdg9$], which includes the 90\degr\ 
inclination angle. 
We therefore can not rule out,
on the $2\,
\sigma$ significance level, the minimum mass of 4.1
\MJ. At the  $2\, \sigma$ level we therefore get 

\begin{equation}
a_1= 1.4^{+1.3}_{-0.8}\,{\rm mas} \ \ ,\,\ M_p=10.1^{+9.5}_{-6.0}\, M_J \ .
\end{equation}

\section{Discussion}

The star \ups\ is the first solar-type star around which a system of
three planet candidates was discovered. Before the recent discovery of
Butler et al.\ (1999) only the pulsar PSR~B1257+12 was known to have
three planets (Wolszczan 1994). Unlike other planet candidates around
main-sequence stars, where the nature of the companions is still in
some doubt (e.g., Black 1997), the multiplicity of the \ups\ system
strongly indicates that the companions are ``proper'' planets. This is
so because we have here one massive object, the parent star, and another
three small objects, with mass ratios of the order of $10^{-2}-10^{-3}$,
orbiting around the large object.  The multiplicity of the small objects
is believed to be one of the key features that characterizes planetary
systems.  Therefore, an estimate of the mass of one of the companions of
\ups\ is important. The analysis of this paper indicates that the mass of
the outermost planet is $10.1^{+4.7}_{-4.6}$ \MJ. This suggests that
planet masses could be substantially larger than 1\,\MJ.

The discussion so far has concentrated on the planet with the longest
orbital period. If the three planets of \ups\ all have the same
inclination, we can estimate the masses of the other two planets too. We
get $1.8\pm 0.8$ and $4.9 \pm 2.3$ \MJ\ for the first and the second
planet, respectively. These estimates {\it assume} that the orbital
planes of all three planets are aligned. However, it is not clear that
this is the case. It is true that the orbital planes of the two outer
planets cannot have very different inclination angles, because a large
relative angle between their planes of motion makes their orbits
dynamically unstable (e.g., Mazeh, Krymolowski \& Rosenfeld 1997;
Holman, Touma \& Tremaine 1997; Krymolowski \& Mazeh 1999).  However,
small angles, like 10\degr -- 20\degr, can not be excluded.

The stability of the planetary system around \ups\ got much attention in
the last few months. A few numerical studies have been performed to find
out if the system is stable on a long timescale (Laughlin \& Adams 1999;
Lissauer 1999; Lissauer \& Rivera 1999; Noyes et al. 1999). All studies
found that the stability strongly depends on the actual masses of the
planets. This is so because the orbits of the second and third planets
put them at a small distance from each other relative to their Hill
radius, a proximity that makes their motion chaotic. The fact that 
the best estimate of this
study for the mass of the third planet is 10\,\MJ, instead
of the 4\,\MJ\ minimum mass, makes the stability of the system even more
questionable. Therefore, our findings necessitate additional careful
surveys of the parameter space of the planetary system, to find
stable enough configurations with 10\,\MJ\ third planet.
Such studies can put a firm upper limit to the mass of the outer planet.

In general, an ideal way to measure a companion mass would be to detect {\it
independently} an astrometric modulation with the same orbital
parameters as those of the spectroscopic solution. We do not yet have
available the astrometric accuracy needed for such a detection. Instead,
we used here the Hipparcos database {\it together} with the
spectroscopic parameters to derive an astrometric orbit.  We are
studying now other planet candidates with our method, finding
interesting results, for 70~Vir in particular. These results are
deferred to a subsequent publication. Mayor reported in IAU Coll.\,170
on a somewhat similar study done by Halbwachs et al. (in preparation),
where they considered the brown-dwarf candidates found in the sample of
nearby K and G stars (Mayor et al. 1997).  Although the details of the
Halbwachs et al. study are not yet published, Mayor reported that the
study found the actual values of sin $i$ of the orbits of their
brown-dwarf companions to be substantially smaller than unity. According
to that study, all the brown-dwarf candidates in their sample
actually have stellar masses.

Obviously, we wait impatiently for the launch of SIM, whose planned
capabilities will allow to detect independently the astrometric orbit of
the {\it two} outermost companions of \ups\ with high S/N.  In the
meantime, while we hold our breath, we need to settle for the Hipparcos
data, which allow only to detect motion induced by planets orbiting with
long periods, of the order of a few years, around nearby main-sequence
stars. As the time base of the radial-velocity measurements is getting
longer, we expect more such planets to be discovered. We therefore
expect in the next few years, even before the new instruments start to
operate, a few more planets whose mass could be estimated.

We wish to express our deep gratitude to the referee, whose extremely
useful comments and wise and thoughtful advice substantially improved
this study. This work was supported by the US-Israel Binational Science
Foundation through grant 97-00460 and the Israeli Science Foundation.

\newpage

\section*{REFERENCES}
\begin{description}

\item Black, D.~C. 1997, ApJL, 490, L171

\item Boss, A.~P. 1996, Nature, 379, 397 

\item Butler, R.~P., Marcy, G.~W., Fischer, D.~A., Brown, T.~W., Contos,
A.~R., Korzennik, S.G., Nisenson, P., \& Noyes, R.~W. 1999,
ApJ, in press

\item Holman, M., Touma, J., \& Tremaine, S. 1997, Nature, 386, 254  

\item Krymolowski, Y., \& Mazeh, T. 1999, MNRAS, 304, 720 

\item Latham, D.~W., Mazeh, T., Stefanik, R.~P., Mayor, M., \& Burki,
G. 1989, Nature, 339, 38

\item ESA 1997, The Hipparcos and Tycho Catalogues, ESA SP-1200

\item Laughlin, G., \& Adams, F.C. 1999, submitted for publication

\item Lisauer, J.J. 1999, Nature, 398, 659

\item Lisauer, J.J., \& Rivera, E. 1999, submitted for publication
 
\item van Leeuwen, F. 1997, Space Sc. Rev., 81, 201

\item van Leeuwen, F., \& Evans, D.~W. 1998, A\&AS, 130, 157

\item Marcy, G.W., \& Butler R.~P. 1998, ARAA, 36, 57

\item Marcy, G.~W., Cochran, W.~D., \& Mayor, M. 2000, in Protostars and
Planets IV, ed.  V. Mannings, A. P. Boss \& S. S. Russell (Tucson:
University of Arizona Press), in press
 
\item Mayor, M., \& Queloz, D. 1995, Nature, 378, 355 

\item Mayor, M., Queloz, D., Udry, S., Halbwachs, J.-L. 1997, in IAU
Coll. 161, Astronomical and Biochemical Origins and Search for Life in
the Universe ed. C. B. Cosmovici, S. Boyer, \& D. Werthimer 
(Bolognia: Editrice Compositori) 313

\item Mazeh, T., Goldberg, D., \& Latham, D. W. 1998, ApJL, 

\item Mazeh T., Krymolowski Y., Rosenfeld G., 1997, ApJL, 477, L103

\item Mazeh, T., Latham, D.~W., \& Stefanik R.~P. 1996, ApJ, 466, 415 

\item Mazeh, T., Mayor, M., \& Latham D.~W. 1996, ApJ, 478, 367

\item Noyes, R.~W., Jha, S., Korzennik, S.~G., Krockenberger, M.,
Nisenson, P., Brown, T.~M., Kennelly, E.~J., \& Horner, S.~D. 1997,
ApJL, 483, L111

\item Noyes, R.~W., Korzennik, S.G., Nisenson, P., Holman, M.J., Contos,
A.~R., \& Brown, T.~W. 1999, BAAS, in press

\item Perryman, M.~A.~C., et al. 1996, A\&A, 310, L21

\item Perryman, M.~A.~C., et al. 1997, A\&A, 323, L49

\item S\"oderhjelm, S. 1999, A\&A, 341, 121

\item Wolszczan, A. 1994, Science, 264, 538

\end{description}

\newpage

\begin{center}
FIGURE LEGENDS
\end{center}

\figcaption{The minimum square-root normalized $\chi^2$ statistics 
as a function of $ a_1$ and $i$. 
The continous line is the $a_1 \times \sin i= 0.56\ {\rm mas}$ constraint.}

\figcaption{The minimum square-root normalized $\chi^2$ statistics as a
function of $i$, given the constraint $a_1 \times \sin i= 0.56\ {\rm
mas}$.}

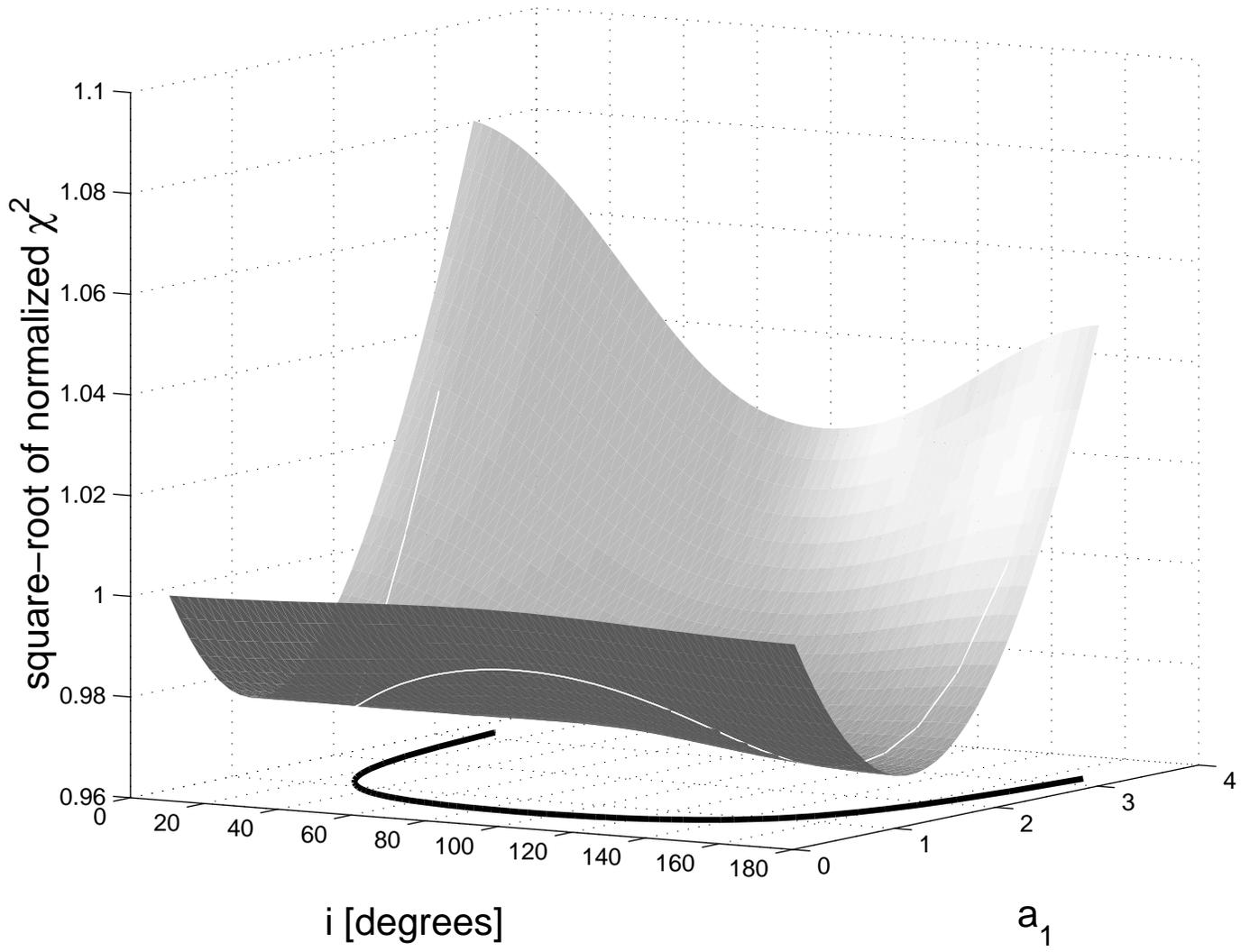
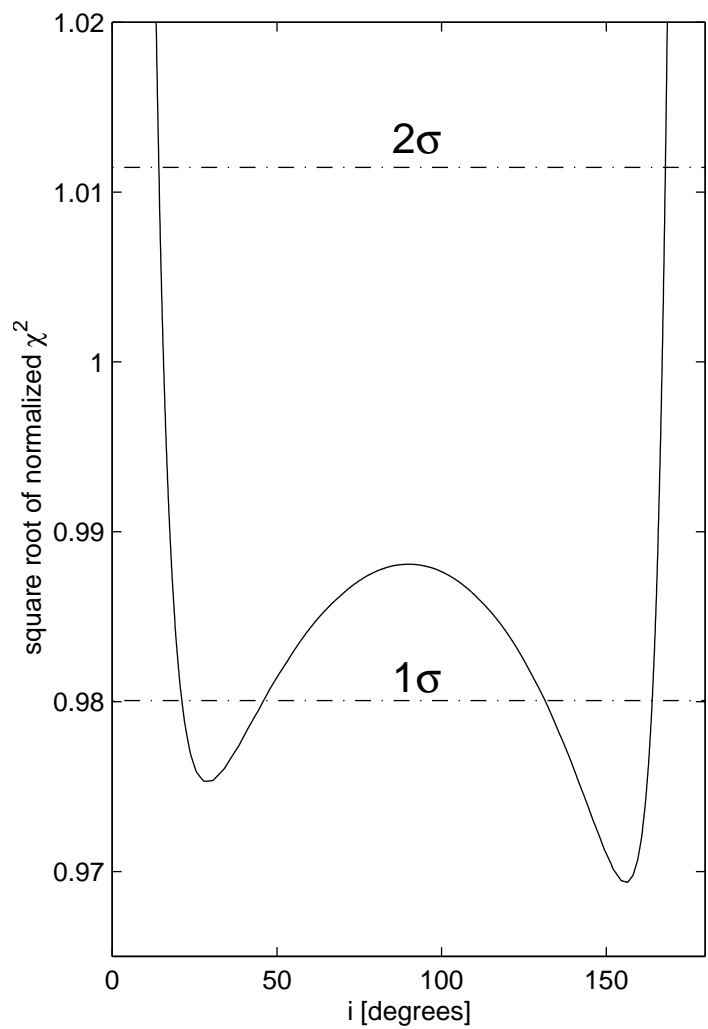
\end{document}